\documentstyle[aps,epsf,overcite]{revtex}
\begin{document}
\newcommand{\beq}{\begin{equation}}
\newcommand{\eeq}{\end{equation}}
\newcommand\bea{\begin{eqnarray}}
\newcommand\eea{\end{eqnarray}}
\def\a{\alpha}
\def\b{\beta}
\def\g{\gamma}
\def\G{\Gamma}
\def\h{\hat}
\def\ch{\chi}
\def\ep{\epsilon}
\def\ph{\phi}
\def\s{\sigma}
\def\th{\theta}
\title{Design of a Protein Potential Energy Landscape by Parameter Optimization}
\author{Julian Lee$^{1,2,3}$, Seung-Yeon Kim$^3$, and Jooyoung Lee$^3$\footnote{Corresponding author: jlee@kias.re.kr}}
\address{$^1$ Department of Bioinformatics and Life Sciences, Soongsil
University, Seoul 156-743, Korea\\
$^2$ Bioinformatics and Molecular Design Technology Innovation
Center, \\
Soongsil University, Seoul 156-743, Korea \\
 $^3$School of Computational Sciences, Korea Institute for Advanced Study,
Seoul 130-722, Korea}
\maketitle
\begin{abstract}
 We propose an automated protocol for designing the energy landscape of a protein energy function
 by optimizing its parameters. The parameters are optimized so that
 not only the global minimum energy conformation becomes native-like, but also the conformations distinct from the
 native structure have higher energies than those close to the native one.
We classify low-energy conformations into three groups, super-native, native-like, and non-native ones.
 The super-native conformations have all backbone dihedral angles fixed to their native values,
 and only their side-chains are minimized with respect to energy.
 On the other hand, the native-like and non-native conformations all correspond to the
 local minima of the energy function.
 These conformations are ranked according to their root-mean-square deviation (RMSD) of backbone coordinates
 from the native structure, and a fixed number of conformations with the lowest RMSD values are defined to be
 native-like conformations, whereas the rest are defined as non-native ones.
 We define two energy gaps $E^{(1)}_{\rm gap}$ and $E^{(2)}_{\rm gap}$.
 The energy gap $E^{(1)}_{\rm gap}$ $(E^{(2)}_{\rm gap})$ is the energy difference between
 the lowest energy of the non-native conformations and the highest energy of the native-like (super-native) ones.
The parameters are modified to decrease both $E^{(1)}_{\rm gap}$ and $E^{(2)}_{\rm gap}$. In addition, the non-native conformations
 with larger values of RMSD are made to have higher energy relative to those with smaller RMSD values.
We successfully apply our protocol to the parameter optimization of the UNRES potential energy,
using the training set of betanova, 1fsd,
 the 36-residue subdomain of chicken villin headpiece (PDB ID 1vii), and the
10-55 residue fragment of staphylococcal protein A (PDB ID 1bdd).
The new protocol of the parameter optimization shows better
performance than earlier methods where only the difference between
the lowest energies of native-like and non-native conformations
was adjusted without considering various degrees of
native-likeness of the conformations. We also perform jackknife
tests on other proteins not included in the training set and
obtain promising results. The results suggest that the parameters
we obtained using the training set of the four proteins are
transferable to other proteins to some extent.
\end{abstract}
\setcounter{equation}{0}

\section{introduction}
The prediction of the three-dimensional structure and the folding
pathway of a protein solely from its amino acid sequence is one of
the most challenging problems in biophysical chemistry. There are
two major approaches to the protein structure prediction, so
called knowledge-based methods and energy-based methods. The
knowledge-based methods\cite{know1,know2,know3,know4}, which
include comparative modeling and fold recognition, use statistical
relationship between sequences and their three-dimensional
structures in the Protein Data Bank (PDB), without deep
understanding of the interactions governing the protein folding.
Therefore, although these methods can be very powerful for
predicting the structure of a protein sequence that has a certain
degree of similarity to those in PDB, they cannot provide the
fundamental understandings of the protein folding mechanism.

  On the other hand, the energy-based methods \cite{ab1,ab2,corn1,corn2,corn3,corn4,corn6},
 which are also called the physics-based methods, are based on
 the thermodynamic hypothesis that proteins adopt native structures that minimize their
 free energies\cite{anfin}. Understanding the fundamental principles of the protein folding
 by these methods will lead not only to the
 successful structure prediction, especially for proteins having no similar sequences in PDB,
 but also to the clarification of the protein folding mechanism.

 However, there have been several major obstacles to the successful application of energy-based methods to the
 protein folding problem. First,
 there are inherent inaccuracies in the potential energy functions which describe the energetics of proteins.
Second, even if the global minimum-energy conformation is
native-like,
 this does not guarantee that a protein will fold into its native structure in a reasonable time-scale
 unless the energy landscape is
 properly designed, as summarized in the Levinthal paradox.

 Physics-based potentials are generally parameterized from quantum mechanical calculations and experimental
  data on model systems\cite{jlee}. However, such calculations and data do not determine the parameters with
  perfect accuracy.
  The residual errors in potential energy functions may have significant effects on simulations of macromolecules such as
  proteins
  where the total energy is the sum of a large number of interaction terms. Moreover, these terms are known to cancel each
   other to a high degree, making their systematic errors even more significant. Thus it is crucial to refine
   the parameters of a potential energy function before it can be successfully applied to the protein folding problem.

 An iterative procedure which systematically refines the parameters of a given
potential energy function was recently proposed\cite{jlee} and
successfully applied to the parameter
optimization\cite{jlee,pill,liwo7,jul} of a UNRES potential
energy\cite{liwo3,liwo4,liwo5}. The method exploits the high
efficiency of the conformational space annealing (CSA)
method\cite{csa1,csa2,csa3,lj,bln} in finding distinct low energy
conformations. For a given set of proteins, whose low-lying local
minimum-energy conformations for a given energy function is found
by the CSA method, one modifies the parameter set so that
native-like conformations of
 these proteins have lower energies than non-native ones. The method consists of the following steps:

\noindent
 (1) Low-lying local minimum-energy conformations are searched with no constraints, which is called the global
CSA search. For many proteins,  the conformations resulting from
the global CSA are non-native conformations for parameters that
are not optimized yet.

\noindent
 (2) Native-like conformations are searched by the local
CSA search, where low-lying local minimum-energy
 conformations are sampled among those whose root-mean-square deviation (RMSD) of the backbone $C^\alpha$ coordinates from the native structure is below
 a given cutoff value $R^{(1)}_{\rm cut}$.

\noindent
 (3) The native-like and non-native conformations from the steps (1) and (2) are added to the structural database
 of each protein.

\noindent
 (4) Among the conformations in the structural database, those with RMSD below a given cutoff value $R^{(2)}_{\rm cut}$ are
 defined as native-like conformations, whereas the rest are defined as non-native ones. The parameters are optimized in
 such a way to minimize the energy gaps
 \begin{equation}
E_{\rm gap} = E^{ \rm N}_{\rm min}-E^{\rm NN}_{\rm min} \label{gaps}
 \end{equation}
for {\it all} proteins in the training set, where $E^{\rm N}_{\rm
min}$ $(E^{\rm NN}_{\rm min})$ is the minimum-energy among the
energies of the native-like (non-native) conformations in the
structural database.

\noindent (5) After the parameters are modified in the step (4),
the conformations in the structural database are not  local
minimum-energy conformations any more. Therefore it is necessary
to reminimize these conformations using the potential energy with
the new parameters.

\noindent (6) In general, with the new parameters, there may exist
many additional low-lying local minima of the potential energy,
which are absent in the structural database. Therefore, it is
necessary to go back to the step (1) and perform CSA searches with
the new parameters.

\noindent These steps are repeated until the performance of the
optimized parameters is satisfactory, i.e., the global CSA search
finds native-like conformations with reasonably small values of
RMSD from their corresponding native structures. Since the size of
the structural database of local minimum-energy conformations
grows after each iteration, the efficiency of the parameter
optimization increases as the algorithm proceeds.

It would be desirable to include many proteins in the training set
that represent many structural classes of proteins. The
optimization method was successfully applied to the parameter
 optimization of the UNRES potential for a training set consisting of three proteins of structural classes
 $\alpha$ and $\alpha / \beta$ \cite{jul},
 without introducing additional multibody terms \cite{corn6,pill,liwo6}.
 However, it was still difficult to optimize the parameters of the UNRES potential for a training set containing
 $\beta$ proteins.


In this work, we propose a new protocol where the parameters are
modified so as to make conformations with larger values of RMSD
have higher values of energy relative to those with smaller values
of RMSD. This goal is achieved by using the following modified
energy \bea E_{\rm modified} = E + 0.3\ {\rm RMSD} \label{modify}
\eea when calculating the energy gaps, where the numerical value
of the coefficient 0.3 is an arbitrarily chosen value. The new
method is more natural than previous methods\cite{jlee,pill,jul}
where non-native conformations were treated equally regardless of
their RMSD values. It also turns out that the new method is much
more efficient than the previous ones and allows us now to
optimize the parameters for a training set containing a $\beta$
protein.
Additional new features are introduced in the current method to
overcome several major drawbacks of the previous methods as below.

First, in previous methods\cite{jlee,pill,jul}, arbitrarily chosen
values of RMSD cutoffs $R^{(1)}_{\rm cut}$ and $R^{(2)}_{\rm cut}$
were used as the criteria
 for separating native-like conformations from non-native ones,
 which were set at each iteration by inspecting the distribution of RMSD values of conformations. This rather arbitrary
 procedure made it difficult to automate the optimization procedure. Moreover, for some proteins,
 the value of $R^{(1)}_{\rm cut}$
 had to be taken as a large number in order to have a non-zero number of native-like conformations,
 in which case the local CSA search is not
meaningful. This can happen for a protein where the initial
parameter set is so bad that there exist no local minimum-energy
conformations which are native-like.
 This problem is solved in the current method by introducing what we call the super-native conformations,
 whose backbone angles are fixed to the values of the native structure and
  only side-chain angles are minimized with respect to the energy. In the current method, the local CSA search
  is defined as the restricted search for the super-native conformations in the space of the side-chain angles.
  Since the $C^\alpha$ RMSD values for
  the super-native conformations are zero by definition,
  an arbitrary cutoff value $R^{(1)}_{\rm cut}$ is no longer necessary.
  Also, the super-native conformations can be found for any parameter
  set. Although the super-native conformations are unstable with
  respect to the energy, minimizing the energy gap between their
  highest energy and the lowest energy of non-native conformations
  has an effect of stabilizing their energies. Therefore,
  due to the reminimization procedure with new optimized parameters, the
  super-native conformations would furnish low-lying local minima with small RMSD values
   which accumulate as the iteration proceeds. This makes the current method
  more efficient than the earlier methods where it was difficult
  to optimize the parameters unless local minimum-energy conformations with
  small values of RMSD exist with the initial parameters.
  In addition to the super-native conformations, we define native-like conformations as the 50 conformations
  with the lowest RMSD values in the structural database. Although 50 is an arbitrary
number, it can be kept as a fixed number, and again the cutoff
value $R^{(2)}_{\rm cut}$ is set automatically. As mentioned
above, the low lying local minima with small RMSD values can be
provided from the reminimization procedure of the super-native
conformations. Generally, the RMSD values of these native-like
conformations become smaller as the iteration of the parameter
optimization continues. Both super-native and native-like
conformations are used for calculating the energy gaps.

Second, we introduce the Linear Programming to systematically
perform the parameter optimization based
 on linear approximation. This allows one to optimize the parameters so that an energy gap of a protein is minimized,
 while {\it imposing the constraints} that the other energy gaps, including those of the other proteins,
 do not increase
 if they are positive,
  and do not become positive if they are negative.  This is in contrast to the
optimization method of earlier works where the protein with the
largest energy gap was selected in turn, and the energy gap of
that protein was reduced
 without imposing any constraint to the energy gaps of the other proteins in the set.
Since the Linear Programming has an effect of simultaneously decreasing
the energy gaps of all the proteins in the training set, it is especially powerful
when there are many proteins in the training set.

  In this work, we successfully apply this method to the optimization of
  linear parameters in the UNRES potential energy,
  for a training set consisting of betanova, 1fsd, the 36-residue subdomain of chicken villin headpiece (PDB ID 1vii),
  and the 10-55 residue fragment of staphylococcal protein A (PDB ID 1bdd).
  We obtain the global minimum energy conformations (GMECs) of these proteins with RMSD values of
  $1.5$, $1.7$, $1.7$, and $1.9$ \AA, respectively.
  The proteins in the training set are $\beta$ (betanova), $\alpha / \beta$ (1fsd), and $\alpha$ (1vii and 1bdd) proteins,
  which cover representative structural classes of small proteins in the nature.
   The basic form of the UNRES potential we use, where the only multibody term is the four-body term,
    is the one that was used for successful prediction of  unknown structures of
proteins in CASP3\cite{corn1,corn4,casp3}. With the optimized
parameters, we have performed jackknife tests on various proteins
not included in the training set, and we find promising results.

\section{Methods}
\subsection{Potential Energy Function}
\begin{figure}
\epsfxsize=15cm
\epsfbox{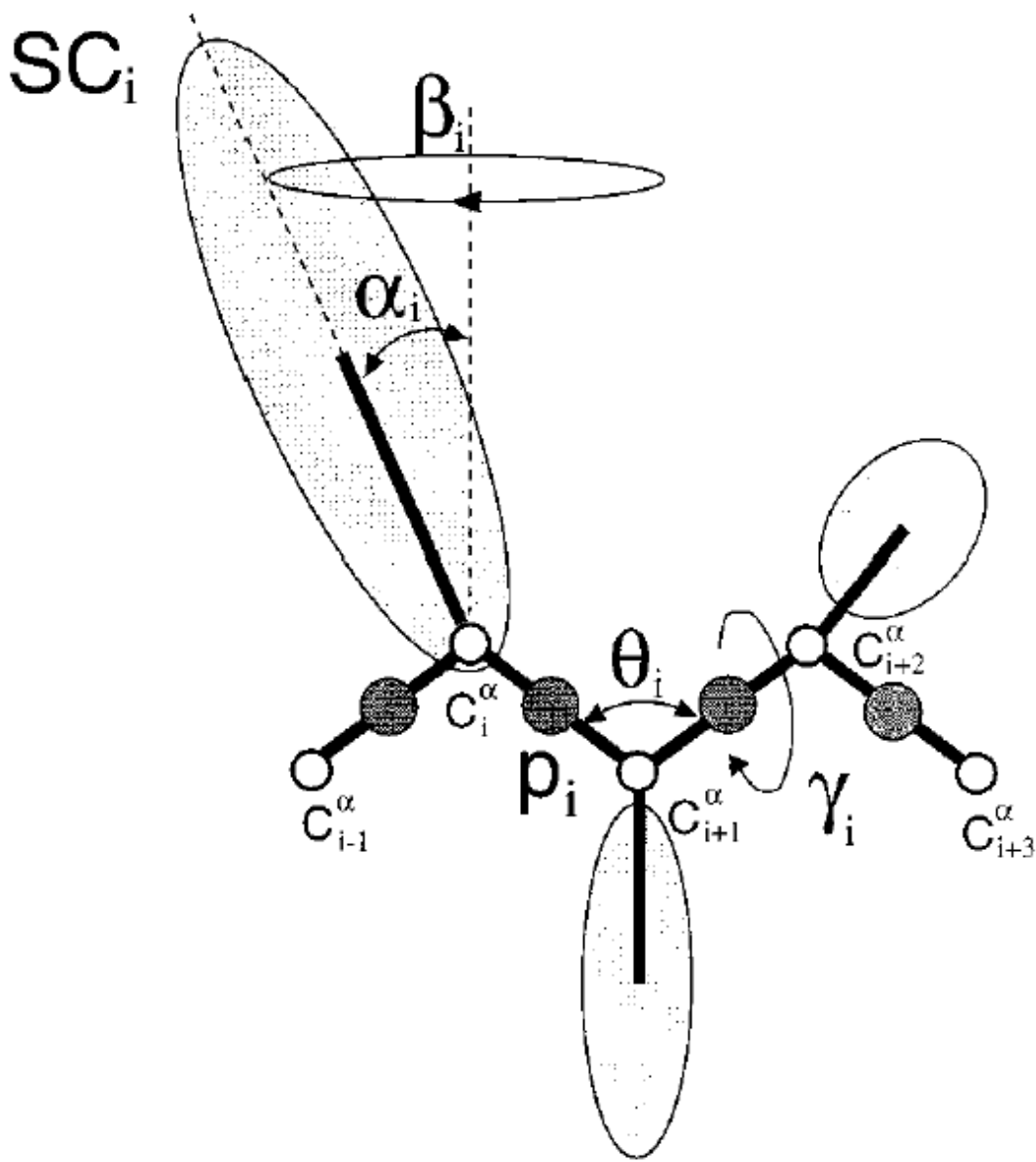} \caption{United-residue
representation of a protein. The interaction sites are side-chain
ellipsoids of different sizes (SC) and peptide-bond centers (p)
indicated by shaded circles, whereas the $\alpha$-carbon atoms
(small empty circles) are introduced to define the backbone-local
interaction sites and to assist in defining the geometry. The
virtual $\rm C^\alpha$-$\rm C^\alpha$ bonds have a fixed length of
3.8 \AA, corresponding to a trans peptide group; the virtual-bond
($\theta$) and dihedral ($\gamma$) angles are variable. Each
side-chain is attached to the corresponding $\alpha$-carbon with a
different but fixed bond length, $b_i$, variable bond angle,
$\alpha_i$, formed by SC$_i$ and the bisector of the angle defined
by C$^\alpha_{i-1}$, C$^\alpha_i$ and C$^\alpha_{i+1}$, and with a
variable dihedral angle $\beta_i$ of counterclockwise rotation
about the bisector, starting from the right side of the
C$^\alpha_{i-1}$, C$^\alpha_i$, C$^\alpha_{i+1}$ frame.}
\label{unres}
\end{figure}

We use the UNRES force field\cite{liwo3,liwo4,liwo5}, where a
polypeptide chain is represented by a sequence of $\alpha$-carbon
($\rm C^\alpha$) atoms linked by virtual bonds with attached
united side-chains (SC) and united peptide groups (p) located in
the middle between the consecutive $\rm C^\alpha$'s (Figure
\ref{unres}). All the virtual bond lengths are fixed: the $\rm
C^\alpha$-$\rm C^\alpha$ distance is taken as $3.8$ \AA, and $\rm
C^\alpha$-$\rm SC$ distances are given for each amino acid type.
The energy of the chain is given by
\begin{eqnarray}
E &=& \sum_{i<j} U_{\rm SCSC}(i,j) +  \sum_{i\neq j}
U_{\rm SCp}(i,j)
     +  \sum_{i<j-1} U_{\rm pp}(i,j) + \sum_i U_{\rm b}(i) \nonumber \\
 & & +  \sum_i U_{\rm tor}(i)+ \sum_i U_{\rm rot}(i)
      + U_{\rm dis} + \sum_{i<j} U^{(4)}_{\rm el-loc}(i,j) \label{pot}
\end{eqnarray}
 As described in detail in the
Appendix of ref.[15], $U_{\rm SCSC}$, $U_{\rm SCp}$, $U_{\rm pp}$,
$U_{\rm tor}$, and $U^{(4)}_{\rm el-loc}$ can be further
decomposed into linear combinations of smaller parts, whose
coefficients are refined in this work.  Here, $U_{\rm SCSC}(i,j)$
represents the mean free energy of the hydrophobic (hydrophilic)
interaction between the side-chains of residues $i$ and $j$, which
is expressed by Lennard-Jones potential, $U_{\rm SCp}(i,j)$
corresponds to the excluded-volume interaction between the
side-chain of residue $i$ and the peptide group of residue $j$,
and the potential $U_{\rm pp}(i,j)$  accounts for the
electrostatic interaction between the peptide groups of residues
$i$ and $j$. The terms $U_{\rm tor}(i)$, $U_{\rm b}(i)$, and
$U_{\rm rot}(i)$ denote the short-range interactions,
corresponding to the energies of virtual dihedral angle torsions,
virtual angle bending, and side-chain rotamers, respectively.
$U_{\rm dis}$ denotes the energy term which forces two cysteine
residues to form a disulfide bridge.  Finally, the four-body
interaction term $U^{(4)}_{\rm el-loc}$ results from the cumulant
expansion of the restricted free energy  of the polypeptide chain.
The functional form eq (\ref{pot}), as well as the initial
parameter set we use, is the one used in the CASP3
exercise\cite{corn1,casp3}. The total number of linear parameters
which we adjust is 715 \cite{param}.

\subsection{Global and Local CSA}
In order to check the performance of a potential energy function
for a given set of parameters, one has to sample super-native,
native-like, and non-native conformations
 for each protein in the training set.
For this purpose we perform two types of conformational search,
the local and global CSA searches. In the local CSA, the backbone
angles of the conformations are fixed to the values of the native
conformations, and only the side-chain energy is minimized with
respect to the energy. We call the resulting conformations the
super-native. The other conformations are obtained from
unrestricted conformational search which we call global CSA. The
conformations obtained from the local and global searches are
added to the structural database of local minimum-energy
conformations for each protein.

\subsection{Parameter Refinement Using Linear Programming}
The changes of energy gaps are estimated by the linear
approximation of the potential energy in terms of parameters.
Among the conformations with non-zero RMSD values in the
structural database, 50 (an arbitrary number) conformations with
the lowest RMSD values are defined as the native-like
conformations, while the rest are defined to be the non-native
ones. Since a potential can be considered to describe the nature
correctly if native-like structures have lower energies than the
non-native ones, the parameters are optimized to minimize the
energy gaps $E^{(1)}_{\rm gap}$ and $E^{(2)}_{\rm gap}$,
\begin{eqnarray}
E^{(1)}_{\rm gap} &=& E^{\rm N} - E^{\rm NN} \nonumber \\
E^{(2)}_{\rm gap} &=& E^{\rm SN} - E^{\rm NN}
\end{eqnarray}
for each protein in the training set, where $E^{\rm N}$ and
$E^{\rm SN}$ are the {\it highest} energies of the native-like and
super-native conformations, respectively, and $E^{\rm NN}$ is the
lowest energy of the non-native conformations. The energies are
the modified ones that are weighted with the RMSD values of the
conformations as in eq (\ref{modify}). Weighting the energy with
the RMSD value has the effect of ``pushing harder" the high RMSD
conformations compared to the ones with lower RMSD values. This
idea is somewhat similar to the hierarchical optimization method
proposed in ref.[16], where the secondary structure contents were
used for the criterion for ranking the nativeness of
conformations. In this work we simply use the RMSD values. The
RMSD value is easier to calculate and consequently it becomes
easier to
 automate the procedure. The parameter
optimization is carried out by minimizing the energy gaps
$E^{(1)}_{\rm gap}$
   and $E^{(2)}_{\rm gap}$ of each
   protein in turn, while imposing the constraints that all the other energy gaps,
   including those from the other proteins, do not increase.

In this work, we adjust only
the linear parameters for simplicity, the total number of them
being 715 for the UNRES potential.
Therefore the energy of a local minimum energy conformation can be written as:
\begin{equation}
E =\sum_j p_j e_j({\bf x_{min}})
\end{equation}
where $e_i$'s are the energy components evaluated with the
coordinates ${\bf x_{min}}$ of a local minimum-energy
conformation. Since the positions of the local minima also depend
on
  the parameters, the full parameter dependence of the energy gaps are nonlinear.
  However, if the parameters are changed by small amounts,
  the energy with the new parameters can be estimated by the linear approximation:

\begin{equation}
\label{Erem} E^{\rm new} \approx E^{\rm old}+ \sum_i (p^{\rm
new}_i-{p}^{\rm old}_i) e_i({\bf x_{min}})
\end{equation}
where the $p^{\rm old}_i$ and $p^{\rm new}_i$ terms represent the
parameters before and after the modification, respectively. The
parameter dependence of the position of the local minimum can be
neglected in the linear approximation, since the derivative in the
conformational space vanishes at local minima\cite{jlee}. The
additional term 0.3 RMSD of eq.(\ref{modify}) vanishes in these
expressions due to the same reason. The change of the energy gaps
are estimated as:
\begin{eqnarray}
\Delta E^{(1)}_{\rm gap} &=& E^{(1)}_{\rm gap}(\{ p^{\rm new}_j \})- E^{(1)}_{\rm
gap}(\{ p^{\rm old}_j \})\nonumber\\
&=&(E^{\rm N}(\{ p^{\rm new}_j \})-E^{\rm NN}(\{ p^{\rm new}_j \}))
-(E^{\rm N}(\{ p^{\rm old}_j \})-E^{\rm NN}(\{ p^{\rm old}_j \}))\nonumber\\
&=& \sum_j (e_j^{\rm N}-e_j^{\rm NN})(p^{\rm new}_j-p_j^{\rm old}) \label{linapp1}
\end{eqnarray}
\begin{eqnarray}
\Delta E^{(2)}_{\rm gap} &=& E^{(2)}_{\rm gap}(\{ p^{\rm new}_j \})- E^{(2)}_{\rm
gap}(\{ p^{\rm old}_j \})\nonumber\\
&=&(E^{\rm SN}(\{ p^{\rm new}_j \})-E^{\rm NN}(\{ p^{\rm new}_j \}))
-(E^{\rm SN}(\{ p^{\rm old}_j \})-E^{\rm NN}(\{ p^{\rm old}_j \}))\nonumber\\
&=& \sum_j (e_j^{\rm SN}-e_j^{\rm NN})(p^{\rm new}_j-p_j^{\rm old}) \label{linapp2}
\end{eqnarray}
The magnitude of the parameter change $\delta p_j \equiv p_j^{\rm
new}-p_j^{\rm old}$ is bounded by a certain fraction $\epsilon$ of
$p^{\rm old}_j$. We use $\epsilon=0.01$ in this study. First, the
vector $\delta p_j$ is chosen within the bound to decrease the
energy gap $\Delta E^{(1)}_{\rm gap}$ of the selected protein as
much as possible while imposing the constraints that any positive
values among $E^{(2)}_{\rm gap}$ and the energy gaps of the other
proteins do not increase and negative values do not become
positive. Denoting the energy gaps of the $k$-th protein as
$E^{(p=1,2)}_{\rm gap}(k)$ and assuming the $i$-th protein is
selected for the decrease of the energy gap, this problem can be
phrased as follows:

\noindent
Minimize \bea \Delta E^{(1)}_{\rm gap}(i) = \sum_j
(e_j^{\rm N}(i)-e_j^{\rm NN}(i))(p^{\rm new}_j-p_j^{\rm old}) \eea
with constraints
\beq
|\delta p_i|  \le \epsilon
\eeq
\beq
\Delta E^{(2)}_{\rm gap}(i) =  \sum_j (e_j^{\rm SN}(i)-e_j^{\rm NN}(i))(p^{\rm new}_j-p_j^{\rm old})
\le \cases{  0   &{\rm if} $E^{(2)}_{\rm gap}(i)>0$ \cr
-E^{(2)}_{\rm gap}(i)  &{\rm otherwise} \cr}
\eeq
\beq
\Delta E^{(p=1,2)}_{\rm gap}(k \neq i) =  \sum_j (e_j^{\rm (S)N}(k)-e_j^{\rm NN}(k))(p^{\rm new}_j-p_j^{\rm old})
\le \cases{  0   &{\rm if} $E^{(p)}_{\rm gap}(k)>0$ \cr
-E^{(p)}_{\rm gap}(k)  &{\rm otherwise} \cr}
\eeq


This is a global optimization problem where the linear parameters $p_j$ are the {\it variables}.
The object function to minimize and the constraints are all linear in $p_j$. This type of
the optimization problem is called the Linear Programming.
It can be solved exactly, and many algorithms have been developed for solving the Linear Programming problem.
We use the primal-dual method with supernodal Cholesky factorization \cite{meszaros} in this work,
which finds a reasonably accurate answer in a short time.

After minimizing $\Delta E^{(1)}_{\rm gap}(i)$, we solve the same
form of linear programming where now $\Delta E^{(2)}_{\rm gap}(i)$
are the objective function and the other energy gap changes become
constrained. Then we select another protein and repeat this
procedure (300 times in this work) of minimizing $\Delta
E^{(1)}_{\rm gap}$ and $\Delta E^{(2)}_{\rm gap}$ in turn.

\begin{figure}
\epsfysize=9cm
\epsfbox{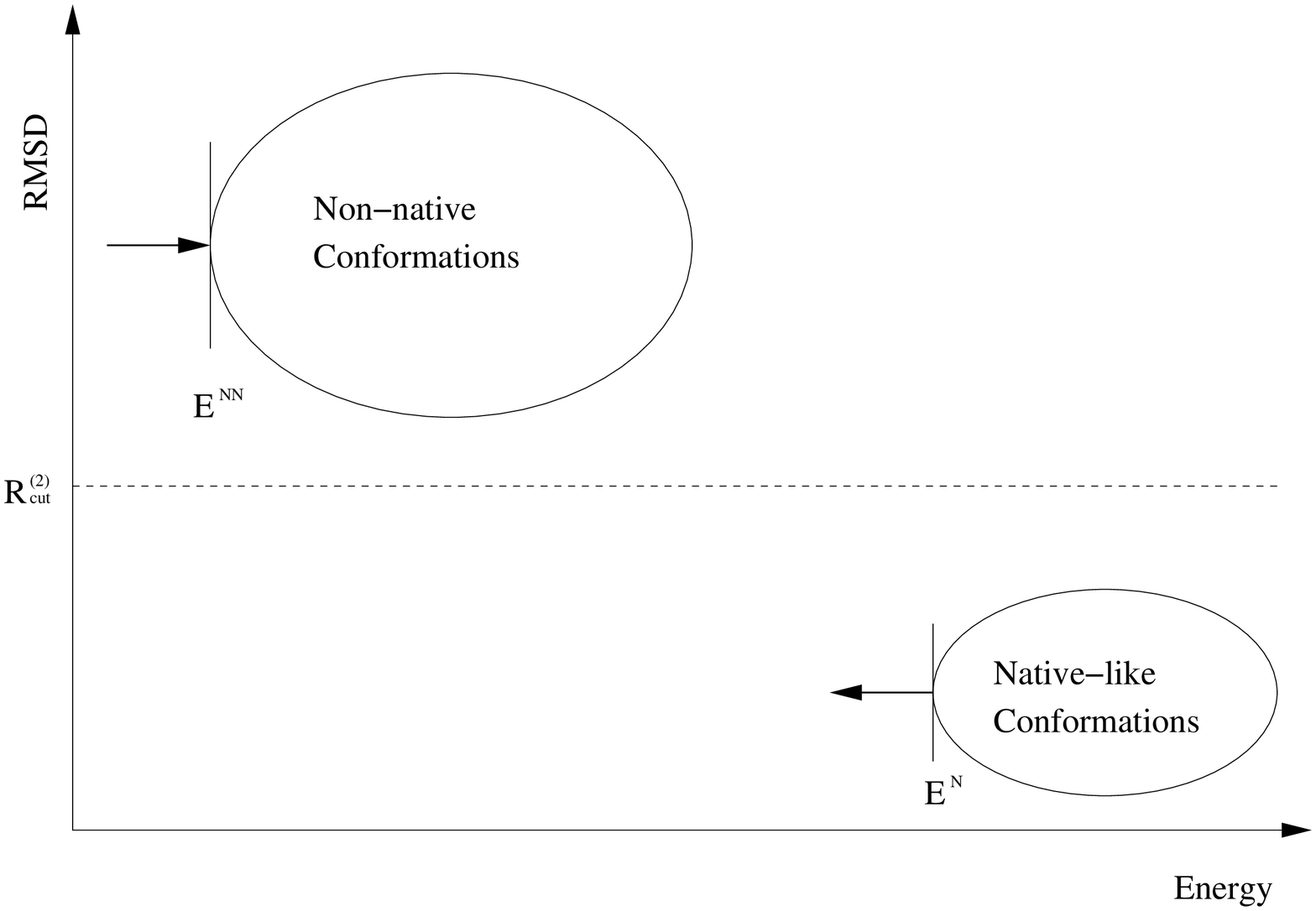}
\caption{The schematic of the old method in terms of the energy and RMSD.
The conformations in the structural database are divided into native-like and non-native
conformations with an arbitrary RMSD cutoff $R^{(2)}_{\rm cut}$.
The minimum energies of of these two families define the energy gap. (See text.).
The arrows indicate the direction of the optimization.}
\label{schema}
\end{figure}

\begin{figure}
\epsfysize=9cm
\epsfbox{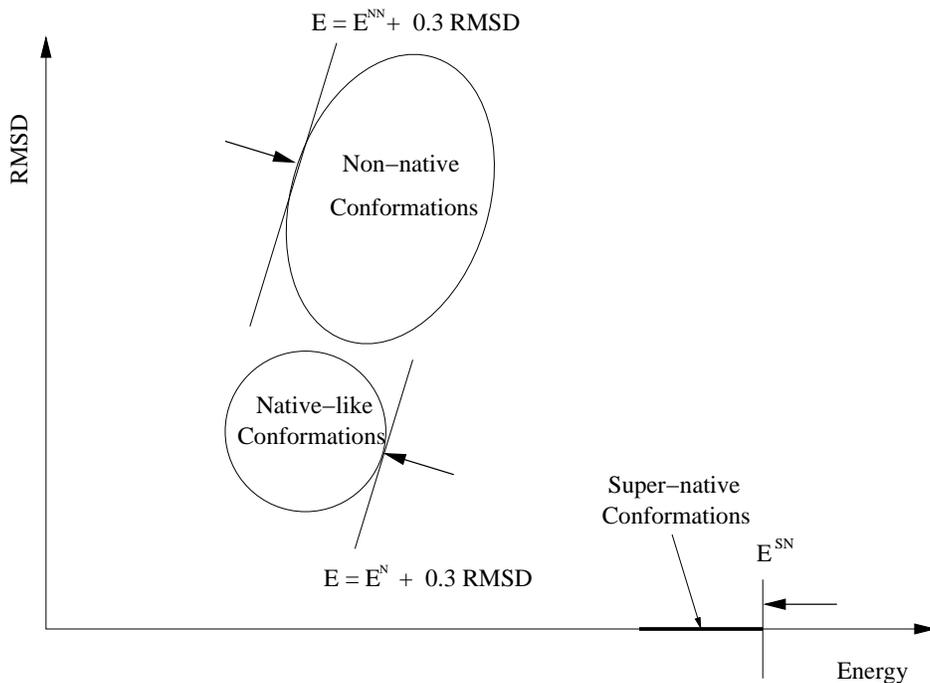}
\caption{The schematic of the new method in terms of the energy
and RMSD. The energy along the vertical axis is the one without
the 0.3 RMSD term. Among the conformations with non-zero RMSD, 50
conformations with the lowest RMSD values are selected as the
native-like conformations, and the rest are considered as the
non-native conformations. The super-native conformations are those
with zero RMSD. The super-native conformations furnish the
candidates for the low-lying native-like local minima after the
reminimization procedure with new optimized parameters.
 The minimum modified energy of the non-native family, and the {\it maximum} modified energies
of native-like and super-native families define the energy gaps (See text.).
The arrows indicate the direction of the optimization.}
\label{schemb}
\end{figure}

The current optimization procedure is different from the one used in the earlier works\cite{jul}
where optimization was performed without
using super-native conformations, and energy was not weighted with RMSD values.
The earlier procedure and the current one are shown schematically in
Figure \ref{schema} and Figure \ref{schemb}, respectively,
in terms of energy and RMSD.

\subsection{Reminimization and New Conformational Search}
 Since the procedure of the previous section was based on
 the linear approximation eqs (\ref{linapp1}) and (\ref{linapp2}),
 we now have to evaluate the true energy gaps
 using the newly obtained parameters. The breakdown of the linear approximation may come
  from two sources.
  First, the conformations corresponding to the local minima of
  the potential
  for the original set of parameters are no longer necessarily so for the new
  parameter set. For this reason, we reminimize the energy of these
  conformations with the new parameters. Since the super-native
  conformations are not local minimum-energy conformations, even
  with the original parameters, the reminimization of these
  conformations with the new parameters would furnish low-lying local
  minima with small values of RMSD.
Second, the local minima obtained using CSA method with the
original parameter set may constitute only a small fraction of
low-lying local minima. After the change of the parameters, some
of the local minima which were not considered due to their
relatively high energies, can now have low energies for the new
parameter set. It is even possible that entirely distinct
low-energy local minima appear. Therefore these new minima are
taken into account by performing subsequent CSA searches (See
section B.) with the newly obtained parameter set.

\subsection{Update of the Structural Database and Iterative Refinement of Parameters}
 The low-lying local energy minima found in the new conformational searches are added into the
  energy-reminimized conformations to form a structural database of local
  energy minima. The conformations in the database are used to obtain the energy gaps,
  which are used for the new round of parameter refinement.
  As the procedure of [CSA $\rightarrow$
  parameter refinement $\rightarrow$ energy reminimization] is repeated,
  the number of conformations in the structural database increases\cite{jul}.
  This  iterative procedure is continued until sufficiently good native-like conformations are
  found from the global CSA search.

\section{Results}
\subsection{Four Proteins in the Training Set}
  We apply our protocol to a training set consisting of four proteins.
  They are the designed protein betanova, 1fsd, 36 residue subdomain of chicken villin headpiece (HP36 or 1vii),
  and 10-55 fragment of the B-domain of staphylococcal protein
  A (1bdd), which are 20, 28, 36, 46 residues long, respectively.
  The protein betanova is a $\beta$ protein, 1fsd is a
  $\alpha/\beta$ protein, and the rest are $\alpha$ proteins,
  which represent structural classes of small
  proteins.
  The initial parameter set is the one used in CASP3\cite{corn1,casp3}.

\begin{figure}
\epsfxsize=10cm
\epsfbox{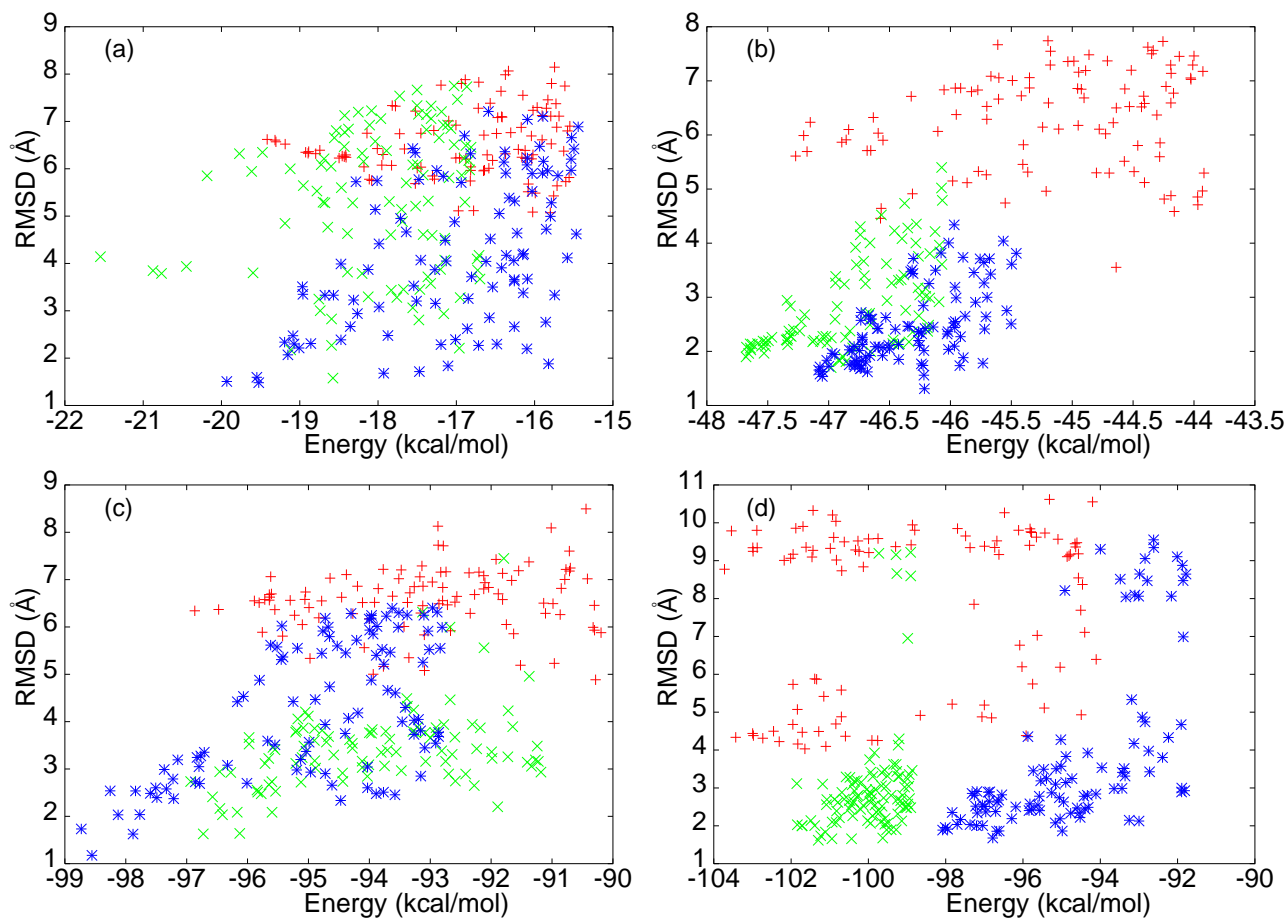}
\caption{Plots of the UNRES energy and ${\rm C^\alpha}$ RMSD (from
the native structure) of four proteins obtained from global CSA
search using the initial and refined parameters. The red, green,
and blue crosses denote the results obtained using the parameters
before the optimization, after the 28-th iteration, and after the
40-th iteration. The results are shown for (a) betanova, (b) 1fsd,
(c) 1vii, and (d) 1bdd.} \label{pro4}
\end{figure}

\begin{figure}
\epsfxsize=10cm
\epsfbox{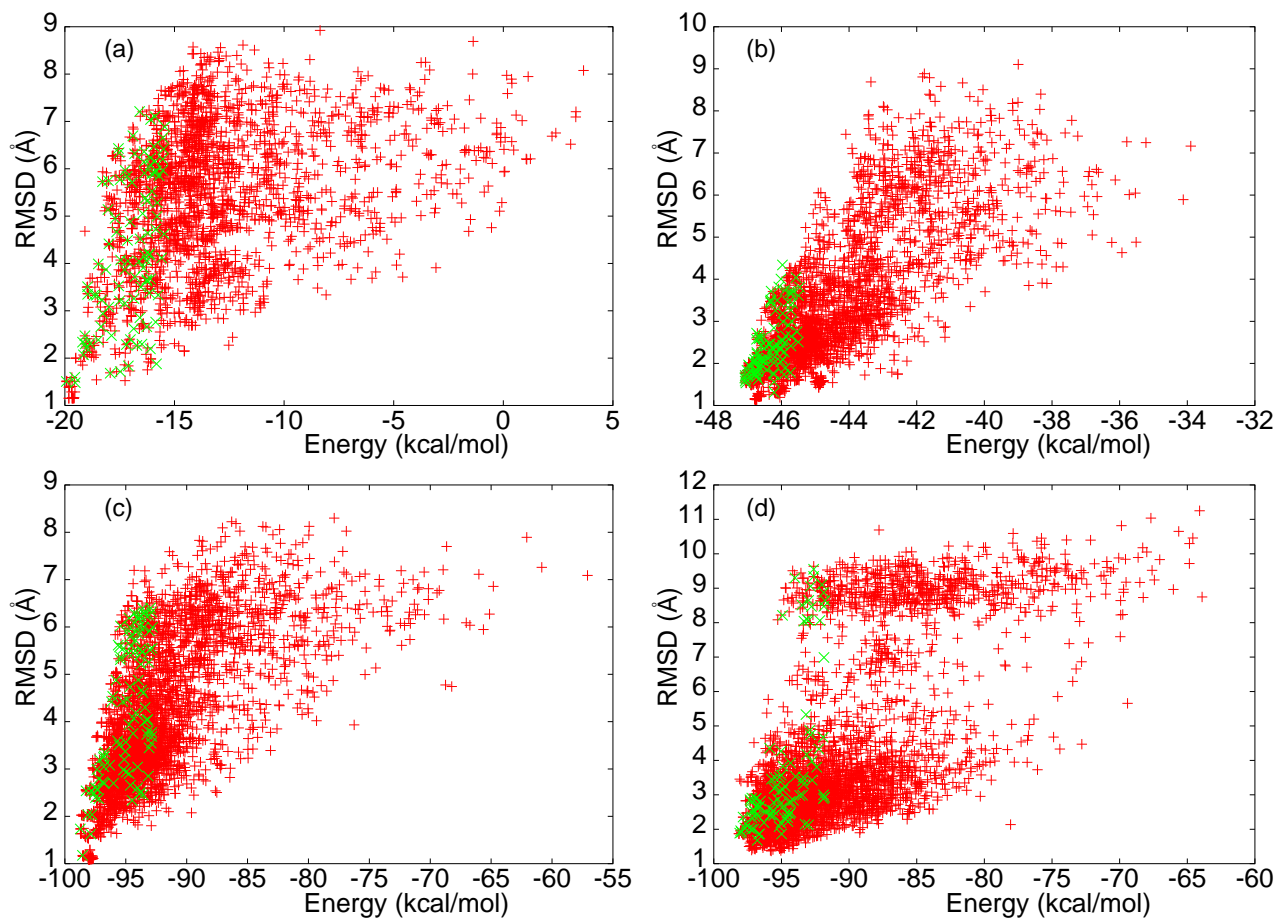}
\caption{Plots of the UNRES energy and ${\rm C^\alpha}$
RMSD (from the native structure) of local-minimum energy conformations
in the structural database of four proteins accumulated after the 40-th iteration
of parameter optimization (red) and the new conformations obtained from the global CSA using these parameters (green).
The results are shown for (a) betanova, (b) 1fsd, (c) 1vii, and (d) 1bdd.}
\label{dbase}
\end{figure}

\begin{figure}
\epsfxsize=18cm
\epsfbox{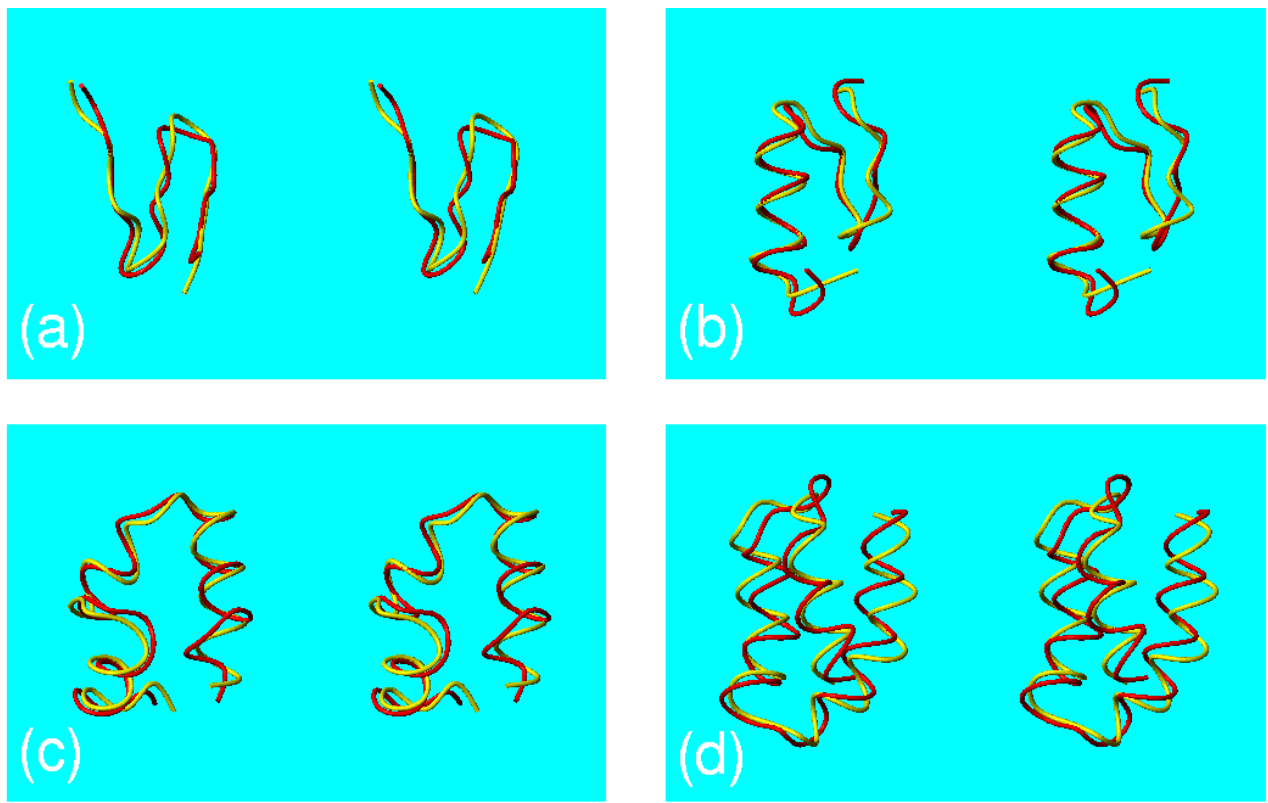}
\caption{The ${\rm C^\alpha}$ trace of GMEC found with the
optimized parameters is shown together with the native structure
for each of the four proteins in the training set. The native
structure is shown in red and the GMEC  is shown in yellow. The
GMECs are shown for (a) betanova, with the RMSD value $1.51$ \AA,
(b) 1fsd, with RMSD value $1.65$ \AA, (c) 1vii, with RMSD value
$1.73$ \AA, and (d) 1bdd, with the RMSD value $1.89$ \AA. The
figures were prepared with the program MOLMOL
\cite{molmol}.}\label{pro4_CA}
\end{figure}

   Fifty conformations were sampled in each CSA search,
and the global minimum-energy conformations (GMECs) found with the
initial parameters have RMSD values of $6.6$, $5.6$, $6.3$ and
$9.5$ \AA, respectively, and the smallest values of RMSD found
from the CSA search are $5.1$, $3.6$, $4.9$ and $4.0$ \AA. After
the 28-th iteration of the parameter refinement, the conformations
with smaller values of RMSD are found from the global CSA search.
The GMECs have RMSD values of $4.1$, $1.9$, $2.7$ and $3.1$ \AA\
and the smallest values of RMSD found are $1.6$, $1.7$, $1.6$ and
$1.6$ \AA. The RMSD values become even smaller after the 40-th
iteration with RMSDs of GMECs being $1.5$, $1.7$, $1.7$ and $1.9$
\AA \ and the smallest values of RMSD being  $1.5$, $1.3$, $1.2$
and $1.7$ \AA.
    The RMSDs of the GMECs and the lowest RMSDs for these
    parameters are shown in the table \ref{rms_res}.
    The results of the global search with the initial and optimized parameter set for the four proteins
    are also plotted in different colors in terms of energy and RMSD in Figure
    \ref{pro4}.
     The local energy conformations accumulated in the structural databases after the 40-th
     iteration of the parameter refinement are shown in Figure \ref{dbase} along with the global CSA search results.
    The $C^\alpha$ traces of the GMECs of the four proteins found using the
    parameters obtained after the 40-th iteration of optimization
are shown in Figure \ref{pro4_CA}  along with the native
conformations.


We also observe a linear slope of $0.3$ in the energy vs. RMSD plot for the low-lying states.
It turns out that the energy landscape designed this way assures a good foldability.
In fact, the direct Monte Carlo folding simulation with the UNRES potential using the parameters
after the 40-th iteration of the refinement, could successfully fold all four proteins into their native states\cite{sykim}.

\subsection{Jackknife Tests}

  We have performed conformational searches for
   proteins not contained in the training set, which are usually called jackknife tests.
   We selected proteins of various structural classes, composed of no more than 60 amino acids residues.
   We considered proteins from NMR experiments,
   since the protein structures in the training set are all determined by NMR spectroscopy. We find
   that the performance of the optimized parameters is reasonably good, and the optimized parameter set
    provides better performance compared to the results from the initial parameter set.
  This implies that the optimized parameters are not
  overfitted to the four proteins in the training set, but are transferable to other proteins to some extent.

\begin{figure}
\epsfxsize=10cm \epsfbox{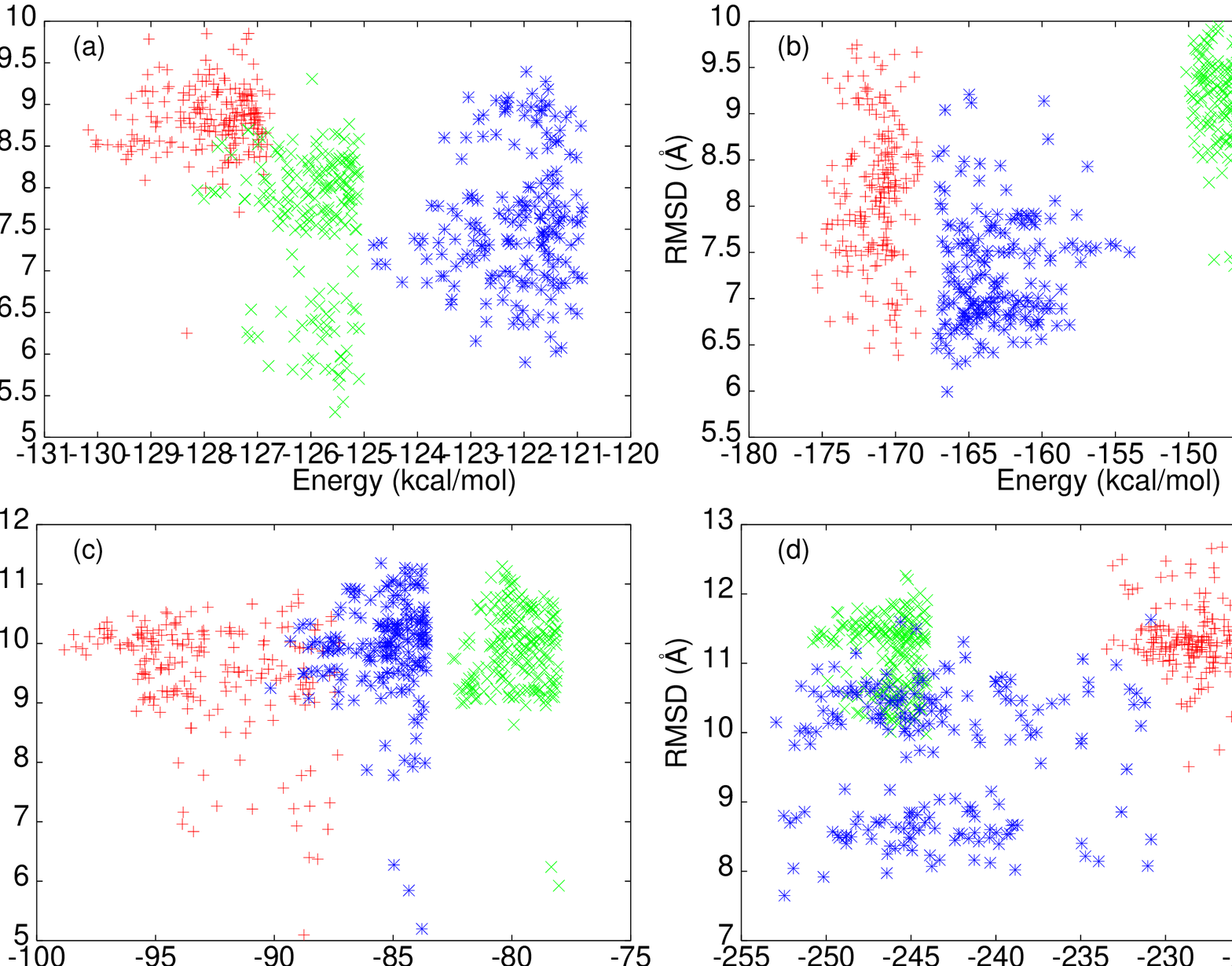} \newpage \epsfxsize=10cm
\epsfbox{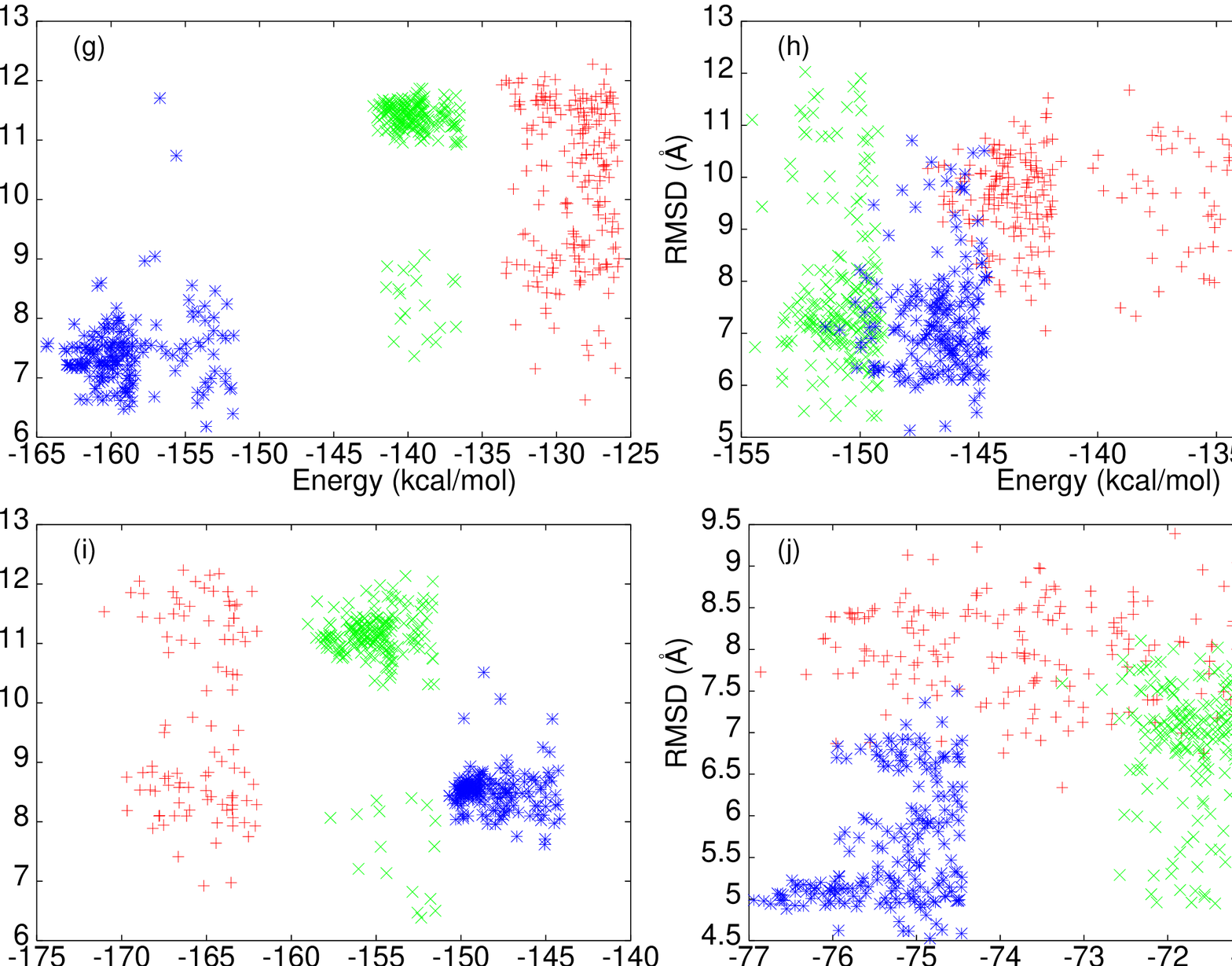}
\caption{The results of the jackknife test. Plots of the UNRES
energy versus ${\rm C^\alpha}$ RMSD (from the native structure) of
the low-lying local enery-minimum conformations are shown.
Conformations obtained from the global CSA using initial
parameters, parameters obtained after the 28-th iteration and
40-th iteration of optimization are shown in red, green, and blue,
respectively. The results are shown for (a) 1bbg, (b) 1ccn, (c)
1hnr, (d) 1kbs, (e) 1neb, and (f) 1bba (g) 1idy, (h) 1prb, (i)
1pru, and (j) 1zdb .} \label{jack}
\end{figure}

\begin{figure}
\epsfysize=20cm
\epsfbox{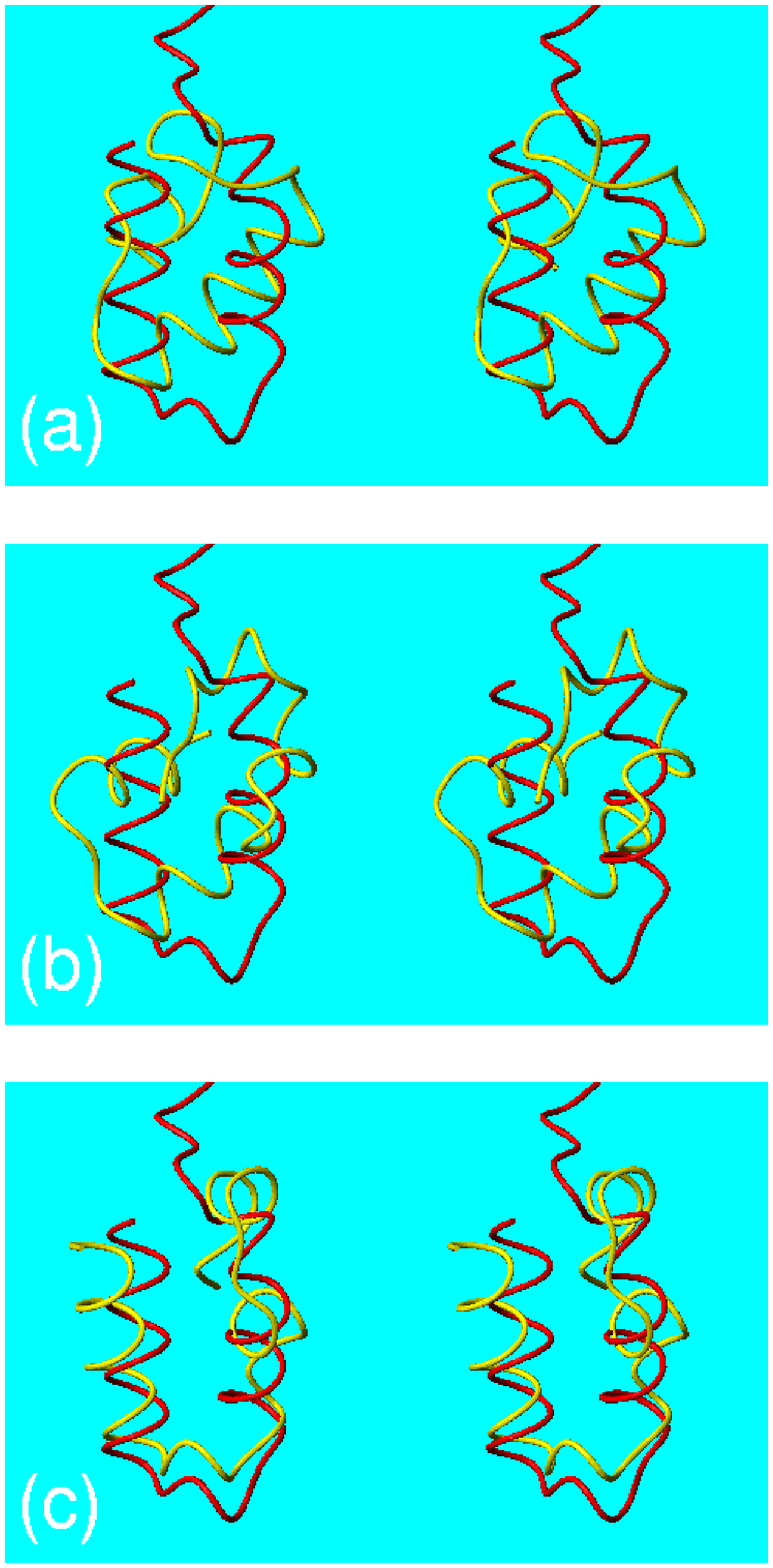}
\caption{(a) The ${\rm C^\alpha}$ trace of the GMECs of 1zdb for
various parameter sets. The native structure is shown in red and
the GMECs found with the optimized parameters are shown in yellow.
It should be noted that there are conformations with even smaller
values of RMSD among those found from the CSA search.
 The results are shown with (a) initial parameters (RMSD 7.7 \AA),
 (b) parameters after the 28-th iteration (RMSD 7.6 \AA), and
 (c) parameters after the 40-th iteration (RMSD 5.0 \AA).
 We observe that the GMEC becomes more and more similar to the native structure as the
 parameter optimization continues, although this protein is not included in the training set.
 This strongly suggests that the optimized parameters are transferable to other proteins.
 The figures are prepared with the program MOLMOL \cite{molmol}.}
 \label{zdb3}
\end{figure}

We have considered proteins 1bbg, 1ccn, 1hnr, 1kbs, 1neb, 1bba,
1idy, 1prb, 1pru, and 1zdb, with the number of amino-acid residues
being 40, 46, 47, 60, 60, 36, 54, 53, 56, and 38, respectively.
The protein 1bbg, 1ccn, and 1hnr are $\alpha / \beta$ proteins,
1kbs and 1neb are $\beta$ proteins, and  the rest are $\alpha$
proteins.
The RMSD values of GMECs and the best structures found with
initial and optimized parameter set are shown in Table
\ref{rms_res}. The results are also shown in terms of energy and
RMSD in Figures \ref{jack}. We find that the results for the
protein 1zdb is particularly notable. In figure \ref{zdb3}, the
$C^\alpha$ traces of the GMECs found with initial and optimized
parameters are shown together with the native structure. We see
that although this protein is not included in the training set,
the GMEC becomes more and more similar to the native structure as
the parameter optimization procedure is continued. These results
suggest that the parameters we obtained from the training set of
the four proteins provide better performances than the initial
parameter set, and are transferable to other proteins.

\section{Discussion}

 We have proposed a general protocol for the force field parameter optimization and landscape design,
 and applied it to the UNRES potential.
 We optimized the 715 linear parameters so that they correctly describe
 the energetics of four proteins simultaneously.
 This optimized parameter set yielded GMECs with RMSD values of $1.5$, $1.7$, $1.7$,
 and $1.9$ \AA\ for betanova, 1fsd, 1vii, and 1bdd, respectively.
 In the process we designed the energy landscape to have a good
 foldability\cite{sykim}. It seems that the
 current parameter optimization method achieves this goal by
 constructing the protein folding funnel\cite{funnel}, which is believed to be
 an essential property of the protein energy functions in nature. It would be interesting to see how many proteins can be
energetically well described using
  a given force field. This should provide a good measure for the efficacy
of existing force fields.

In contrast to the earlier protocols\cite{jlee,jul}, where the
value of RMSD cutoff values was specified for each protein at each
iteration to define native-like conformations, we now defined 50
conformations with the lowest RMSD values
 in the structural database for each protein as the native-like
 conformations
 and used super-native structures,
 which have zero RMSD values
 by definition, to furnish candidates for low-lying native-like
 conformations of small values of RMSD.
  This enabled us to automate the whole procedure using a shell script.
 However, there is still some arbitrariness in our protocol,
 such as choosing 50 native-like conformations,
 and giving the slope of $0.3$ in eq (\ref{modify}).
 We also tried the values of $0.1$ and $0.5$, with similar results as in the case of $0.3$.
 We will have to devise a way of choosing the optimal value of the slope.
 Finally, it should be noted that although we have considered only the UNRES potential for parameter
optimization in this work, it is straightforward to apply the same
procedure to other potentials such as ECEPP\cite{ecepp4},
AMBER\cite{amber}, CHARMM\cite{charmm} with various solvation
terms\cite{ooi, wessen}. All these points are left for the future
study.


\acknowledgements{We thank Ki Hyung Joo and Il-Soo Kim for useful
discussions and helps in fulfilling this work. This work was
supported by grant No. R01-2003-000-11595-0 (Jooyoung Lee) and No.
R01-2003-000-10199-0 (Julian Lee) from the Basic Research Program
of the Korea Science \& Engineering Foundation. The calculation
was carried out on Linux PC cluster of 214 AMD processors at
KIAS.}

\newpage
\begin{table}
\caption{The RMSD values of the GMECs found from the global CSA
search using the initial parameters, and the optimized parameters
after 28-th and 40-th iterations (units in \AA). The numbers
inside the parentheses are the smallest values of RMSD found. The
structural class and the chain length of each protein is also
shown inside the parenthesis next to the protein name. }
\begin{tabular}{|c||c|c|c|}
protein &  0-th & 28-th & 40-th  \\
\hline
betanova ($\beta$ : 20 aa) & 6.6 (5.1) & 4.1 (1.6) & 1.5 (1.5) \\
\hline
1fsd ($\alpha/\beta$ : 28 aa) & 5.6 (3.6) & 1.9 (1.7) & 1.7 (1.3) \\
\hline
1vii ($\alpha$ : 36 aa) & 6.3 (4.9) & 2.7 (1.6) & 1.7 (1.2) \\
\hline
1bdd ($\alpha$ : 46 aa) & 9.6 (4.0) & 3.1 (1.6) & 1.9 (1.7) \\
\hline \hline
1bbg ($\alpha/\beta$ : 40 aa) & 8.7 (6.3) & 7.9 (5.3) & 7.3 (5.9) \\
\hline
1ccn ($\alpha/\beta$ : 46 aa) & 7.7 (6.4) & 9.5 (7.0) & 6.5 (6.0) \\
\hline
1hnr ($\alpha/\beta$ : 47 aa) & 9.9 (5.1) & 9.7 (5.9) & 9.2 (5.2) \\
\hline
1kbs ($\beta$ : 60 aa) & 11.2 (9.5) & 11.3 (10.0) & 10.1 (7.6) \\
\hline
1neb ($\beta$ : 60 aa) & 10.9 (9.3) & 11.3 (8.8) & 9.6 (9.1) \\
\hline
1bba ($\alpha$ : 36 aa) & 8.9 (8.1) & 8.1 (6.8) & 12.0 (10.7) \\
\hline
1idy ($\alpha$ : 54 aa) & 11.9 (6.6) & 11.6 (7.4) & 7.5 (6.2) \\
\hline
1prb ($\alpha$ : 53 aa) & 10.2 (7.0) & 11.1 (5.4) & 7.1 (5.1) \\
\hline
1pru ($\alpha$ : 56 aa) & 8.4 (7.1) & 11.3 (6.4) & 8.4 (7.6) \\
\hline
1zdb ($\alpha$ : 38 aa) & 7.7 (6.3) & 7.6 (4.9) & 5.0 (4.5) \\
\end{tabular}
\label{rms_res}
\end{table}

\begin{references}
\bibitem{param} The number of parameters are larger than 709 used in ref \cite{jul}
since now we allow different weights to the four-body terms for
parallel and anti-parallel strands. Also, whereas the side-chain
peptide group interaction and 14 rescaling (See Appendix.) were
performed in an asymmetric manner in earlier works, we do it
symmetrically.
\end{references}
\end{document}